\documentclass[prd,aps,nofootinbib,showpacs,showkeys]{revtex4}

\def\XXint#1#2#3{{\setbox0=\hbox{$#1{#2#3}{\int}$}
     \vcenter{\hbox{$#2#3$}}\kern-.5\wd0}}

\usepackage{graphicx}
\usepackage{bm}
\usepackage{amssymb}
\usepackage{amsmath}
\usepackage{euscript}

\begin{document}

\title{Quantum Fluctuations of Coulomb Potential as a
Source of Flicker Noise. \\ The Influence of External Electric
Field}

\author{Kirill~A.~Kazakov}\email{kirill_kazakov@comtv.ru}

\affiliation{Department of Theoretical Physics,
Physics Faculty,\\
Moscow State University, $119899$, Moscow, Russian Federation}

\begin{abstract}
Fluctuations of the electromagnetic field produced by quantized
matter in external electric field are investigated. A general
expression for the power spectrum of fluctuations is derived
within the long-range expansion. It is found that in the whole
measured frequency band, the power spectrum of fluctuations
exhibits an inverse frequency dependence. A general argument is
given showing that for all practically relevant values of the
electric field, the power spectrum of induced fluctuations is
proportional to the field strength squared. As an illustration,
the power spectrum is calculated explicitly using the kinetic
model with the relaxation-type collision term. Finally, it is
shown that the magnitude of fluctuations produced by a sample
generally has a Gaussian distribution around its mean value, and
its dependence on the sample geometry is determined. In
particular, it is demonstrated that for geometrically similar
samples,  the power spectrum is inversely proportional to the
sample volume. Application of the obtained results to the problem
of flicker noise is discussed.
\end{abstract}

\maketitle

\section{Introduction}\label{introduction}

As is well-known, power spectra of voltage fluctuations in all
conducting materials exhibit a universal profile in the
low-frequency limit, which is close to inverse frequency
dependence. Fluctuations characterized by the power spectrum of
this type are called usually $1/f,$ or flicker, noise. Although
this noise dominates only at low frequencies, experiments show the
presence of the $1/f$-component in the whole measured frequency
band up to $10^6 Hz.$ Despite numerous attempts, no lower
frequency bound for the $1/f$ law has been found. In addition to
that, it is generally accepted that $1/f$-noise produced by a
sample is universally characterized by the following properties 1)
it is (roughly) inversely proportional to the sample volume, 2) it
is Gaussian, and 3) its part induced by external electric field is
proportional to the field strength squared.

A number of mechanisms has been suggested to explain the origin of
$1/f$-noise \cite{buck}. There is a widespread opinion that this
noise arises from resistance fluctuations, which is quite natural
taking into account the property 3) mentioned above. Indeed, for a
given current through the sample, the mean squares of voltage and
resistance fluctuations are proportional to each other, the
current squared being the proportionality coefficient. It has been
proposed that the resistance fluctuations possessing the other
properties of flicker noise might result from temperature
fluctuations \cite{voss}, fluctuations in the carrier mobility
\cite{hooge,klein} or in the number of carriers caused by surface
traps \cite{mcwhorter}. All these models, however, have restricted
validity, because they involve one or another assumption specific
to the problem under consideration. For instance, assuming that
the resistance fluctuations arise from the temperature
fluctuations, one has to choose an appropriate spatial correlation
of these fluctuations in order to obtain the desired profile of
the power spectrum. Similarly, the model of Ref.~\cite{mcwhorter}
requires specific distribution of trapping times. In addition to
that, the models proposed so far reproduce the $1/f$-profile only
in a restricted range of frequencies, require an appropriate
normalization of the power spectrum, etc. At the same time,
ubiquity of flicker noise and universality of its properties
suggest existence of a simple and universal, and therefore,
fundamental origin. It is natural to look for this reason in the
quantum properties of charge carriers. In this direction, the
problem has been extensively investigated by Handel and co-workers
\cite{handel}. Handel's approach is based on the theory of
infrared radiative corrections in quantum electrodynamics. Handel
showed that the $1/f$ power spectrum of photons emitted in any
scattering process can be derived from the well-know property of
bremsstrahlung, namely, from the infrared divergence of the
cross-section considered as a function of the energy loss. Thus,
this theory treats the $1/f$-noise as a relativistic effect (in
fact, the noise level in this theory $\sim \alpha
(\Delta\bm{v})^2/c^2,$ where $\alpha$ is the fine structure
constant, $\Delta\bm{v}$ velocity change of the particle being
scattered, and $c$ the speed of light). It should be mentioned,
however, that the Handel's theory has been severely criticized in
many respects \cite{tremblay,kampen}.

In Refs.~\cite{kazakov1,kazakov}, the role of quantum effects is
considered from a purely nonrelativistic point of view. In
Ref.~\cite{kazakov1}, quantum fluctuations of the electromagnetic
field produced by elementary particles are investigated, and it is
shown, in particular, that the correlation function of the
fluctuations exhibits an inverse frequency dependence in the
low-frequency limit. This result was applied in
Ref.~\cite{kazakov} to the calculation of the power spectrum of
electromagnetic fluctuations produced by a sample. It was proved,
in particular, that the power spectrum possesses the properties
1), 2) of flicker noise, mentioned above. As to the property 3),
it was argued in Ref.~\cite{kazakov} that this requirement is also
met. The argument was based on the assumption of analyticity of
the electron density matrix with respect to the external electric
field. However, this assumption is not valid in general. Thus, the
issue concerning the influence of external field is left open.

The purpose of the present paper is to investigate the influence
of the external electric field on quantum electromagnetic
fluctuations in detail. We will show that for all practically
relevant values of the field strength, the power spectrum of
induced fluctuations is proportional to the field strength squared
indeed.

Inclusion of an external field lowers the system symmetry,
therefore, our first problem below will be to generalize the
results obtained in \cite{kazakov} for spherically-symmetric
systems to systems with axial symmetry. This is done in
Sec.~\ref{calcul}. First of all, we prove in Sec.~\ref{connected}
that the low-frequency asymptotic of the connected part of
correlation function is logarithmic. This result was proved
already in \cite{kazakov}. Although the proof does not rely on the
system symmetries, we give an independent and more simple and
accurate proof of this important fact. Contribution of the
disconnected part is calculated in Sec.~\ref{discon}, and is found
to exhibits an inverse frequency dependence, thus dominating in
the low-frequency limit. Because of the lower system symmetry,
this calculation is much more complicated than in the case
considered in \cite{kazakov}. The obtained expression for the
power spectrum is analyzed in Sec.~\ref{external} where a general
argument is given showing that the field-induced noise is
quadratic in the field strength, which is then illustrated using
the simplest kinetic model with the relaxation-type collision
term. Section \ref{conclude} summarizes the results of the work
and states the conclusion.

\section{Preliminaries}\label{prelim}

Let us consider electromagnetic field produced by a classical
resting particle with mass $m$ and electric charge $e.$ It is
described by the Coulomb potential
\begin{eqnarray}\label{coulomb}&&
A_0 = \frac{e}{4\pi r}\,, \qquad \bm{A} = 0\,.
\end{eqnarray}
\noindent In quantum theory, this form of the electromagnetic
potential is reproduced by the mean fields $\langle {\rm in}|
\hat{A}_0|{\rm in}\rangle,$ $\langle {\rm in}|\hat{\bm{A}}|{\rm
in} \rangle$ calculated far away from the region of particle
localization. If the 3-vector of the mean particle position is
denoted by $\bm{x}_0,$ and that of the point of observation by
$\bm{x},$ then the latter condition means that $|\bm{x} -
\bm{x}_0| \gg D,$ where $D$ is a characteristic length of the
particle wave packet spatial spreading (for instance, the variance
of the particle coordinates). As a result of the quantum evolution
according to the Schrodinger equation, $D$ increases in time, thus
leading to a dispersion of the electromagnetic field produced by
the particle. Furthermore, because of the quantum indeterminacy in
the particle position, the field fluctuates. The correlation
function of the fluctuations is conventionally defined by
\begin{eqnarray}\label{corr}&&
C_{\mu\nu}(x,x') = \frac{1}{2}\langle {\rm
in}|\hat{A}_{\mu}(x)\hat{A}_{\nu}(x') +
\hat{A}_{\nu}(x')\hat{A}_{\mu}(x) |{\rm in}\rangle - \langle {\rm
in}|\hat{A}_{\mu}(x)|{\rm in}\rangle\langle {\rm
in}|\hat{A}_{\nu}(x')|{\rm in}\rangle\,,
\end{eqnarray}
\noindent where $x$ and $x'$ are the spacetime coordinates of two
observation points. Of course, this function is dispersed, too.
Our aim below will be to investigate low-frequency properties of
this dispersion. It is clear that the condition $|\bm{x} -
\bm{x}_0| \gg D$ is irrelevant in this investigation, because the
low-frequency asymptotic of the power spectrum of correlations is
determined largely by the late-time behavior of the function
$C_{\mu\nu}(x,x'),$ where $D$ is unbounded (for a free particle
state, and for large times $t,$ $D$ is a linear function of $t$).

As in Ref.~\cite{kazakov}, we will work within the long-range
expansion of the correlation function, which is a convenient tool
for extracting the leading term of the correlation function. Let
us briefly recall the reasons justifying application of this
expansion. The function $C_{\mu\nu}(x,x')$ can be represented as a
power series in the ratios $l_c/R$ and $D/R,$ where $l_c =
\hbar/mc$ is the Compton length, and $R$ is either $r = |\bm{x} -
\bm{x}_0|$ or $r' = |\bm{x}' - \bm{x}_0|.$ However, as mentioned
above, the ratio $D/R$ cannot be considered small as long as one
is concerned with the low-frequency behavior of correlations. We
overcome this problem by going over to the momentum space, and
work with an expansion in powers of $l_c$ and $\bm{p}/\tilde{D},$
where $\bm{p}$ is the 3-momentum transfer to the particle, and
$\tilde{D} = \sqrt{\langle\bm{q}^2\rangle}$ is the variance of the
particle momentum. Unlike $D,$ the quantity $\tilde{D}$ is time
independent (for free particle states), so the expansion is valid
for all times. By the order of magnitude, the relevant values of
the momentum transfer $|\bm{p}| \sim \hbar/r,$ and therefore,
validity of the expansion in powers of $\bm{p}/\tilde{D}$ requires
only that $$r \tilde{D}\gg \hbar\,, \quad r'\tilde{D}\gg
\hbar\,.$$ All subsequent considerations are carried out under
these conditions.

We recall also that, as was shown in Ref.~\cite{kazakov1}, the
leading term of the correlation function is of zeroth order in
$l_c,$ hence, in the units $\hbar=c=1$ used from now on, it can be
identified as the limit of the correlation function for $m\to
\infty.$ It should be emphasized that this identification is only
formal, in particular, it does not mean that the results obtained
below apply only to heavy particles.

According to Eq.~(\ref{corr}), in order to find the correlation
function of the electromagnetic fluctuations, one has to calculate
the in-in expectation values of the field operators
$\hat{A}_{\mu}$ as well as of their products. In
Refs.~\cite{kazakov1,kazakov}, this was done using the
Schwinger-Keldysh formalism. In particular, it was proved in
Ref.~\cite{kazakov} that the leading low-frequency term of the
correlation function is contained entirely in its disconnected
part [the second term in Eq.~(\ref{corr})]. The connected part of
the correlation function was taken in Ref.~\cite{kazakov} in the
nonsymmetric form $\langle {\rm
in}|\hat{A}_{\mu}(x)\hat{A}_{\nu}(x')|{\rm in}\rangle.$ Although
the proof given there can be carried over to the present case, we
will give an independent and more accurate proof of this important
fact, which avoids complications of the Schwinger-Keldysh method.

Note first of all, that for all values of $t,t',$ the connected
part of the correlation function can be rewritten as
\begin{eqnarray}\label{corr1}&&
\frac{1}{2}\langle {\rm in}|\hat{A}_{\mu}(x)\hat{A}_{\nu}(x') +
\hat{A}_{\nu}(x')\hat{A}_{\mu}(x) |{\rm in}\rangle  =
\frac{1}{2}\langle {\rm in}|T
\{\hat{A}_{\mu}(x)\hat{A}_{\nu}(x')\} +
\tilde{T}\{\hat{A}_{\mu}(x)\hat{A}_{\nu}(x')\}|{\rm in}\rangle\,,
\nonumber
\end{eqnarray}
\noindent where the operation of time ordering $T$ ($\tilde{T}$)
arranges the factors so that the time arguments decrease
(increase) from left to right. Furthermore, for a one-particle
state under stationary external conditions, the state vector
$|{\rm in}\rangle$ can be substituted by the vector $|{\rm out}
\rangle,$ up to a phase factor. In the tree approximation, this
factor is equal to unity, therefore, one can write, taking into
account that $\hat{A}_{\mu}$ is Hermitian,
\begin{eqnarray}\label{corr2}&&
\frac{1}{2}\langle {\rm in}|\hat{A}_{\mu}(x)\hat{A}_{\nu}(x') +
\hat{A}_{\nu}(x')\hat{A}_{\mu}(x) |{\rm in}\rangle  =
\frac{1}{2}\langle {\rm out}|T
\{\hat{A}_{\mu}(x)\hat{A}_{\nu}(x')\}|{\rm in}\rangle
\nonumber\\&& + \frac{1}{2}\langle {\rm
in}|\tilde{T}\{\hat{A}_{\mu}(x)\hat{A}_{\nu}(x')\}|{\rm
out}\rangle = {\rm Re}\langle {\rm out}|T
\{\hat{A}_{\mu}(x)\hat{A}_{\nu}(x')\}|{\rm in}\rangle\,.
\end{eqnarray}
\noindent The latter quantity can be calculated by applying the
usual Feynman rules.

\begin{figure}
\includegraphics{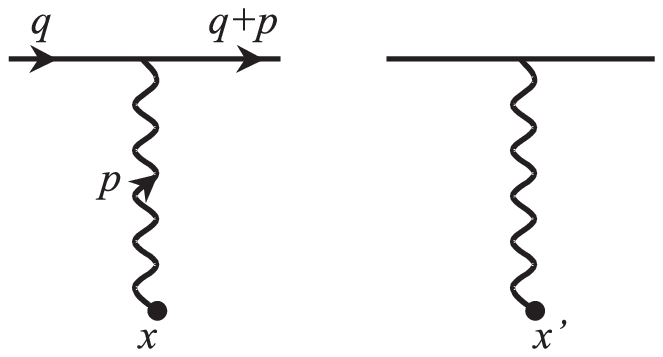}
\caption{Feynman diagrams representing the disconnected part of
correlation function [the second term in Eq.~(\ref{corr})]. Wavy
lines denote the photon propagators, solid lines the massive
particle. $q$ and $p$ are the particle 4-momentum and 4-momentum
transfer, respectively.} \label{fig1}
\end{figure}

Let us assume, for simplicity, that the field-producing particle
is described by a complex scalar $\phi.$ It is not difficult to
show actually that the results derived below are valid for
particles of any spin. This is because in the long-range limit,
the value of the electromagnetic current is fixed by the standard
normalization conditions for the one-particle state, which are
universal for all particle species. Let the gradient invariance be
fixed by the Lorentz condition\footnote{The proof of the
gauge-independence of the leading contribution, given in
\cite{kazakov1}, does not rely on the symmetry properties of the
particle wave function, and hence carries over to the present
case.}
\begin{eqnarray}\label{gauge}
G\equiv\partial^{\mu}A_{\mu} = 0\,.
\end{eqnarray}
\noindent Then the action of the system takes the form
\begin{eqnarray}\label{action}
S[\Phi] &=& S_0[\Phi] + S_{\rm gf}[\Phi]\,, \nonumber\\
S_0[\Phi] &=& {\displaystyle\int} d^4 x
\left\{(\partial_{\mu}\phi^* + i e A_{\mu}\phi^*)
(\partial^{\mu}\phi - ie A^{\mu}\phi) - m^2 \phi^*\phi\right\} -
\frac{1}{4}{\displaystyle\int} d^4 x F_{\mu\nu} F^{\mu\nu}\,,
\nonumber\\ S_{\rm gf}[\Phi] &=& -\frac{1}{2}{\displaystyle\int}
d^4 x~G^2\,,\quad F_{\mu\nu} =
\partial_{\mu} A_{\nu} - \partial_{\nu} A_{\mu}\,,
\end{eqnarray}
\noindent where the Feynman weighting of the gauge condition is
assumed.  The tree diagrams generated by this action, which
contribute to the right hand side of Eq.~(\ref{corr}), are
depicted in Figs.~\ref{fig1}, \ref{fig2}.

To complete this section, let us define the power spectrum
function of fluctuations. We are concerned with correlations in
the values of the electromagnetic fields measured at two distinct
time instants (spatial separation between the observation points,
$|\bm{x} - \bm{x}'|,$ is also kept arbitrary). Accordingly, fixing
one of the time arguments, say, $t',$ we define the power spectrum
function as the Fourier transform of $C_{\mu\nu}(x,x')$ with
respect to $(t - t')$
\begin{eqnarray}\label{corrf}&&
C_{\mu\nu}(\bm{x},\bm{x}',t',\omega) =
\int\limits_{-\infty}^{+\infty}dt C_{\mu\nu}(x,x')e^{i\omega(t -
t')}\,.
\end{eqnarray}
\noindent

\section{Evaluation of the leading contribution}\label{calcul}

Evaluation of the low-frequency asymptotic of the correlation
function proceeds in two steps. First, we will prove in
Sec.~\ref{connected} that the low-frequency asymptotic of the
connected part of $C_{\mu\nu}$ [the first term in
Eq.~(\ref{corr})] is logarithmic. The contribution of the
disconnected part will be calculated in Sec.~\ref{discon}. It will
be shown that this contribution exhibits an inverse frequency
dependence, and thus dominates in the low-frequency limit.

\subsection{Low-frequency asymptotic of the connected part of
correlation function}\label{connected}

\begin{figure}
\includegraphics{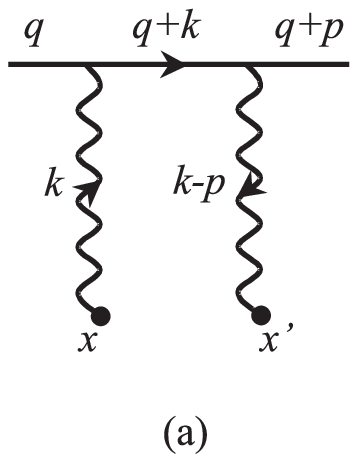}
\includegraphics{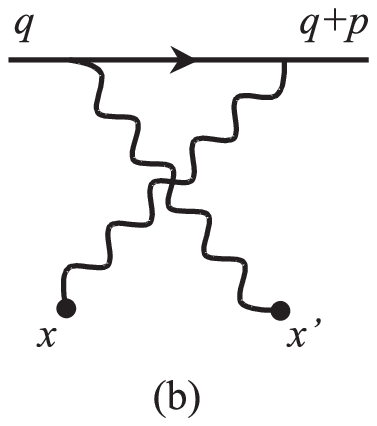}
\includegraphics{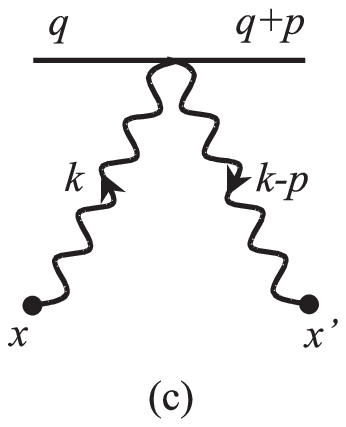}
\caption{Feynman diagrams representing the connected part of
correlation function [the first term in Eq.~(\ref{corr})].}
\label{fig2}
\end{figure}

Before going into detailed calculations, let us first exclude the
diagrams in Fig.~\ref{fig2}, which do not contain the $\hbar^0$
contribution. It is not difficult to see that these are the
diagrams without internal matter lines, i.e. \ref{fig2}(c) in the
present case. Indeed, this diagram is proportional to the integral
$$\int d^4 k
\frac{e^{ik(x-x')}}{k^2(k-p)^2}\ ,$$ which does not involve the
particle mass at all. Taking into account that each external
matter line gives rise to the factor
$(2\epsilon_{\bm{q}})^{-1/2},$ where $\epsilon_{\bm{q}} =
\sqrt{m^2 + \bm{q}^2}\approx m\,,$ we see that the contribution of
diagram \ref{fig2}(c) is proportional to $1/m.$ Hence, on
dimensional grounds, this diagram is proportional to $\hbar.$

The contribution of diagrams \ref{fig2}(a), \ref{fig2}(b) has the
form
\begin{eqnarray}\label{diagen}
B_{\mu\nu} &=& I_{\mu\nu}(x,x') + I_{\mu\nu}(x',x)\,, \nonumber\\
I_{\mu\nu}(x,x') &=& ie^2\iint d^4 z d^4 z'
D^0(x,z)\left[\phi_0^*(z)
\stackrel{\leftrightarrow}{\partial_{\mu}}
D(z,z')\stackrel{\leftrightarrow}{\partial_{\nu}^{\,\prime}}
\phi_0(z')\right]D^0(z',x')\,,
\end{eqnarray}
\noindent where
\begin{eqnarray}
\varphi\stackrel{\leftrightarrow}{\partial_{\mu}}\psi &=&
\varphi\partial_{\mu}\psi - \psi\partial_{\mu}\varphi\,,
\nonumber\\
D(x,y) &=& \int\frac{d^4 k}{(2\pi)^4}\frac{e^{-ik(x-y)}}{m^2 - k^2
- i0}\,, \nonumber\\
D^{0} &\equiv& D|_{m=0}\,,
\end{eqnarray} and $\phi_0$ is the given particle state.
\noindent Introducing the Fourier transform of $I_{\mu\nu}(x,x')$
$$\tilde{I}_{\mu\nu}(\bm{x},\bm{x}',t',\omega) =
\int\limits_{-\infty}^{+\infty}d t
I_{\mu\nu}(x,x')e^{i\omega(t-t')}\,,$$ and going over to the
momentum space, one finds
\begin{eqnarray}\label{diagenk1}
\tilde{I}_{\mu\nu}(\bm{x},\bm{x}',t',\omega) &=& e^2\iint
\frac{d^3 \bm{q}}{(2\pi)^3} \frac{d^3
\bm{p}}{(2\pi)^3}\frac{a^*(\bm{q})a(\bm{q} +
\bm{p})}{\sqrt{2\epsilon_{\bm q}2\epsilon_{{\bm q} + \bm{p}}}}
e^{-ip^0(t' - t_0) + i\bm{p}\bm{x}'}
\tilde{J}_{\mu\nu}(p,q,\bm{x}-\bm{x}',\omega)\,, \\
p^0 &=& \epsilon_{\bm{q} + \bm{p}} - \epsilon_{\bm{q}}\,,
\nonumber
\end{eqnarray}
\noindent where
\begin{eqnarray}\label{diagenk3}
\tilde{J}_{\mu\nu}(p,q,\bm{x}-\bm{x}',\omega) = && - i\int
\frac{d^3 \bm{k}}{(2\pi)^3} e^{i\bm{k}(\bm{x}-\bm{x}')}(2q_{\mu} +
k_{\mu})(2q_{\nu} + k_{\nu} + p_{\nu})\nonumber\\&&\times
\left.D^0(k)D(q+k)D^0(k-p)\right|_{k^0=\omega}\,.
\end{eqnarray}
\noindent Here $q_{\mu}$ is the particle 4-momentum, and
$a(\bm{q})$ its momentum wave function at some time instant $t_0.$
The function $a(\bm{q})$ is normalized by
\begin{eqnarray}\label{norm}
\int\frac{d^3 \bm{q}}{(2\pi)^3}|a(\bm{q})|^2 = 1\,,
\end{eqnarray}\noindent and is generally of the form
\begin{eqnarray}\label{abrel}
a(\bm{q}) = b(\bm{q}) e^{-i\bm{q}\bm{x}_0}\,,
\end{eqnarray}\noindent
where $\bm{x}_0$ is the particle mean position, and $b(\bm{q})$
describes the momentum space profile of the particle wave packet.

Let us now show that the low-frequency asymptotic of
$\tilde{J}_{\mu\nu}$ is logarithmic. We note, first of all, that
in the long-range limit, the 4-momenta $k_{\mu},p_{\mu}$ in the
vertex factors can be neglected in comparison with $q_{\mu},$
because the leading contribution comes from integration over small
$k_{\mu}.$ By the same reason, the factor
$\exp\{i\bm{k}(\bm{x}-\bm{x}')\}$ can be set equal to unity. Next,
introducing the Schwinger parameterization of the propagators, we
rewrite $\tilde{J}_{\mu\nu}$ as
\begin{eqnarray}&&
\tilde{J}_{\mu\nu}(p,q,\omega) = 4q_{\mu}q_{\nu}\int \frac{d^3
\bm{k}}{(2\pi)^3}\exp\left\{i\left(x[\omega^2 - \bm{k}^2] +
y[\omega^2 - \bm{k}^2 + 2m\omega - 2(\bm{k}\bm{q})]
\right.\right.\nonumber\\&&\left.\left. + z[(\omega - p^0)^2 -
(\bm{k} - \bm{p})^2] \right)\right\} =
\frac{q_{\mu}q_{\nu}}{2\pi^3}\iiint_{0}^{\infty}d x d y d z
\left(\frac{\pi
e^{-i\pi/2}}{x+y+z}\right)^{3/2}\exp\left\{i\frac{(y\bm{q} -
z\bm{p})^2}{x+y+z}\right\}\nonumber\\&& \times \exp\left\{i\left(
x\omega^2 + y[\omega^2 + 2m\omega] + z[(\omega - p^0)^2 - \bm{p}^2
]\right)\right\}\,. \nonumber
\end{eqnarray}\noindent Changing the integration variables $y\to xy,$ $z\to
xz,$ and integrating over $x$ gives
\begin{eqnarray}&&
\tilde{J}_{\mu\nu}(p,q,\omega) =
\frac{q_{\mu}q_{\nu}}{4\pi}\iint_{0}^{\infty}d y d z
\left\{(y\bm{q} - z\bm{p})^2 \right.\nonumber\\&&\left. + (1 + y +
z)\left(\omega^2 + y[\omega^2 + 2m\omega] + z[(\omega - p^0)^2 -
\bm{p}^2 ]\right)\right\}^{-3/2}\,. \nonumber
\end{eqnarray}\noindent The singularity of the latter integral at
$\omega = 0$ comes from integration over small $y,z,$ therefore,
it is the same as the singularity of the integral
\begin{eqnarray}&&
\tilde{J}'_{\mu\nu}(p,q,\omega) \nonumber\\&& =
\frac{q_{\mu}q_{\nu}}{4\pi}\iint_{0}^{\infty}d y d z
\left\{y^2\bm{q}^2 - 2yz(\bm{q}\bm{p}) + \omega^2 + y[\omega^2 +
2m\omega] + z[(\omega - p^0)^2 - \bm{p}^2 ]\right\}^{-3/2}
\nonumber\\&& =
\frac{q_{\mu}q_{\nu}}{2\pi}\int_{0}^{\infty}\frac{d y}{(\omega -
p^0)^2 - \bm{p}^2 - 2y(\bm{q}\bm{p})} \left\{y^2\bm{q}^2 +
\omega^2 + y[\omega^2 + 2m\omega] \right\}^{-1/2} \nonumber\\&& =
\frac{q_{\mu}q_{\nu}}{2\pi}\int_{0}^{\infty}\frac{d y}{(\omega -
p^0)^2 - \bm{p}^2 - 2\omega y(\bm{q}\bm{p})} \left\{y^2\bm{q}^2 +
1 + y[\omega + 2m] \right\}^{-1/2} \,. \nonumber
\end{eqnarray}\noindent After the transformations performed, the singularity
of the last integral for $\omega \to 0$ reappears at $y\to
\infty,$ and hence, it coincides with the singularity of the
integral
\begin{eqnarray}&&
\tilde{J}''_{\mu\nu}(p,q,\omega) = \frac{q_{\mu}q_{\nu}}{2\pi
|\bm{q}|}\int^{\infty}\frac{d y}{y}\frac{1}{(\omega - p^0)^2 -
\bm{p}^2 - 2\omega y(\bm{q}\bm{p})} \nonumber\\&& =
\frac{q_{\mu}q_{\nu}}{2\pi |\bm{q}|\{(\omega - p^0)^2 -
\bm{p}^2\}}\ln\left.\frac{y}{(\omega - p^0)^2 - \bm{p}^2 - 2\omega
y(\bm{q}\bm{p})}\right|^{\infty} \sim \frac{q_{\mu}q_{\nu}}{2\pi
|\bm{q}|p^2}\ln\frac{1}{\omega}\,.
\end{eqnarray}\noindent It is not difficult to verify that the
obtained expression agrees with the results of Sec.~III~A of
Ref.~\cite{kazakov}. Thus, we have proved that the connected part
of the correlation function diverges for $\omega \to 0$ only
logarithmically.

\subsection{Low-frequency asymptotic of the disconnected part of
correlation function}\label{discon}

Let us turn to the disconnected part of the correlation function.
To find its Fourier transform, we have to evaluate the integral
\begin{eqnarray}\label{meanfourier}
\int\limits_{-\infty}^{+\infty}dt\, \langle {\rm
in}|A_{\mu}(x)|{\rm in}\rangle e^{i\omega(t-t')} =
\frac{eq_{\mu}}{\epsilon_{\bm{q}}}e^{-i\omega
t'}\tilde{I}(\bm{r},\omega)\,,
\end{eqnarray}\noindent
where
\begin{eqnarray}
\tilde{I}(\bm{r},\omega) &=& \int\limits_{-\infty}^{+\infty}dt
\left\{ e^{i\omega t}
\iint\frac{d^3\bm{q}}{(2\pi)^3}\frac{d^3\bm{p}}{(2\pi)^3}
\frac{e^{i\bm{p}\bm{r}}}{\bm{p}^2}e^{-ip^0(t-t_0)}b^*(\bm{q})
b(\bm{q} + \bm{p})\right\}\,, \quad \bm{r} = \bm{x} -
\bm{x}_0\,.\nonumber
\end{eqnarray}\noindent
To the leading order of the long-range expansion, $b(\bm{q} +
\bm{p})$ can be substituted here by $b(\bm{q}).$ Using also
$$p^0 \approx \frac{(\bm{p} + \bm{q})^2}{2m} -
\frac{\bm{q}^2}{2m}$$ gives
\begin{eqnarray}
\tilde{I}(\bm{r},\omega) = 2\pi e^{i\omega t_0}
\iint\frac{d^3\bm{q}}{(2\pi)^3}\frac{d^3\bm{p}}{(2\pi)^3}
\delta\left(\omega - \frac{\bm{p}^2 + 2 \bm{pq}}{2m}\right)
\frac{e^{i\bm{p}\bm{r}}}{\bm{p}^2}|b(\bm{q})|^2\,. \nonumber
\end{eqnarray}\noindent
The function $b(\bm{q})$ will be assumed to possess the symmetry
of the external field. Thus, in the presence of a homogeneous
external electric field, the function $b(\bm{q})$ is
axially-symmetric; taking $z$ axis in the direction of the field,
one has $b(\bm{q}) = b(q_{\bot},q_z),$ where $q_{\bot} =
\sqrt{\bm{q}^2 - q^2_z}$ is the transverse component of the
particle momentum. In this case, $\tilde{I}$ is a function of
$\zeta = r_z$ and $\rho = \sqrt{\bm{r}^2 - \zeta^2},$
$\tilde{I}(\bm{r},\omega) = \tilde{I}(\rho,\zeta,\omega),$ and
hence, averaging over transverse directions, one can write
\begin{eqnarray}\label{fourier3}
\tilde{I}(\rho,\zeta,\omega) &=& 2\pi e^{i\omega t_0}
\iint\frac{d^3\bm{q}}{(2\pi)^3}\frac{d^3\bm{p}}{(2\pi)^3}
\delta\left(\omega - \frac{\bm{p}^2 + 2 \bm{pq}}{2m}\right)
\frac{J_0(|\bm{p}|\rho\sin\phi)}{\bm{p}^2}e^{i|\bm{p}|\zeta\cos\phi}
|b(q_{\bot},q_z)|^2\,, \nonumber
\end{eqnarray}\noindent where $\phi$ is the
angle between $\bm{p}$ and $z$ axis, and $J_0$ is the Bessel
function. To evaluate the $\bm{p}$ integral it is convenient to
introduce a spherical coordinate system, with the polar axis
pointing in the direction of the vector $\bm{q}.$ Let the
azimuthal and polar angles of $\bm{p}$ in this system be denoted
by $\varphi,\theta,$ and those of $z$ axis by $\varphi',\theta',$
respectively. Then
$$\cos\phi = \sin\theta\sin\theta'\cos(\varphi-\varphi') +
\cos\theta\cos\theta'\,,$$ and
\begin{eqnarray}\label{fourier31}
\tilde{I}(\rho,\zeta,\omega) &=& \frac{me^{i\omega t_0}}{(2\pi)^2}
\int\frac{d^3\bm{q}}{(2\pi)^3}|\beta(q_{\bot},q_z)|^2
\int\limits_{0}^{\pi}d\theta\sin\theta\int\limits_{0}^{2\pi}d\varphi
\frac{J_0(u\rho\sin\phi)e^{iu\zeta\cos\phi}} {\sqrt{\bm{q}^2
\cos^2\theta + 2m\omega}}\,, \\ u &=& - |\bm{q}|\cos\theta +
\sqrt{\bm{q}^2 \cos^2\theta + 2m\omega}\,,\nonumber
\end{eqnarray}\noindent where it is assumed that $\omega >0.$
Since the integrand in this formula
depends on the difference $\varphi-\varphi',$ the result of
integration over $\varphi$ is independent of $\varphi'.$ Setting
the latter equal to zero, and taking $\phi$ as the integration
variable yields
\begin{eqnarray}\label{fourier32}
\EuScript{I}\equiv\int\limits_{0}^{2\pi}d\varphi
J_0(u\rho\sin\phi)e^{iu\zeta\cos\phi} = 2\int\limits_{\theta -
\theta'}^{\theta + \theta'}d\phi \frac{\sin\phi\,
J_0(u\rho\sin\phi)e^{iu\zeta\cos\phi}}{\sqrt{[\cos\phi -
\cos(\theta + \theta')][\cos(\theta - \theta') - \cos\phi]}}\,.
\end{eqnarray}\noindent
It is seen from Eq.~(\ref{fourier3}) that the singularity at
$\omega = 0$ comes from integration over $\theta\approx \pi/2.$
Without changing the singular contribution, therefore, one can set
$\theta = \pi/2$ in the integral (\ref{fourier32}), which after
simple transformations takes the form
\begin{eqnarray}\label{fourier33}
\EuScript{I} = 4\int\limits_{0}^{1}d v
\frac{J_0\left(u\rho\sqrt{1-\sin^2\theta'v^2}
\right)\cos\left(uv\zeta\sin\theta'\right)}{\sqrt{1-v^2}}\,.
\end{eqnarray}\noindent
In the long-range limit, the argument of the Bessel function is
large, so one can use the asymptotic formula $$J_0(z) =
\sqrt{\frac{2}{\pi z}}\cos\left(z - \frac{\pi}{4}\right)\,.$$
Thus, the above integral involves rapidly oscillating
trigonometric functions, and the leading contribution comes from
integration around the point of stationary phase of the integrand.
Assuming $\zeta >0,$ and decomposing the product of cosines into a
sum, one sees that the range of integration contains one such
point, if $\zeta < r\sin\theta',$ namely
$$v_0 = \frac{\zeta}{r\sin\theta'}\,,$$ and hence
\begin{eqnarray}\label{fourier34}
\EuScript{I} &\approx& 2\sqrt{\frac{2}{\pi}}\ {\rm
Re}\int\limits_{0}^{1}d v
\frac{\displaystyle\exp\left\{i\left(u\rho\sqrt{1-\sin^2\theta'v^2}
+ uv\zeta\sin\theta' - \frac{\pi}{4}\right)\right\}}
{\sqrt{(1-v^2)u\rho\sqrt{1-\sin^2\theta'v^2}}} \nonumber\\
&\approx& 2\sqrt{\frac{2}{\pi}}\ {\rm Re}\int\limits_{0}^{1}d v
\frac{\displaystyle\exp\left\{i\left(u r -
\frac{ur^3}{2\rho^2}\sin^2\theta'(v-v_0)^2 -
\frac{\pi}{4}\right)\right\}}
{\sqrt{(1-v^2)u\rho\sqrt{1-\sin^2\theta'v^2}}} \nonumber\\
&\approx& 2\sqrt{\frac{2}{\pi}}\ {\rm Re}
\frac{\displaystyle\exp\left\{i\left(u r -
\frac{\pi}{2}\right)\right\}}
{\sqrt{(1-v_0^2)u\rho\sqrt{1-\sin^2\theta'v_0^2}}}
\sqrt{\frac{2\pi\rho^2}{ur^3\sin^2\theta'}} =
\frac{4\sin(ur)}{\displaystyle ur\sqrt{\sin^2\theta'-
\frac{\zeta^2}{r^2}}} \,.
\end{eqnarray}\noindent Substituting this in Eq.~(\ref{fourier3}),
introducing a new spherical system of coordinates, with the polar
axis in $z$ direction, and noting that $\theta'$ is the polar
angle of the vector $\bm{q}$ gives
\begin{eqnarray}\label{fourier35}
\tilde{I}(\rho,\zeta,\omega) &=& \frac{me^{i\omega t_0}}{4\pi^4 r}
\int\limits_{\arcsin\zeta/r}^{\pi-\arcsin\zeta/r}
\frac{d\theta'\sin\theta'}{\displaystyle\sqrt{\sin^2\theta'-
\frac{\zeta^2}{r^2}}}\int\limits_{0}^{\infty}d q\,
q^2|\beta(q,\theta')|^2 \int\limits_{0}^{\pi}d\theta
\frac{\sin\theta\sin(ur)}{\displaystyle u \sqrt{q^2 \cos^2\theta +
2m\omega}}\,, \nonumber
\\ u &=& - q\cos\theta + \sqrt{q^2 \cos^2\theta +
2m\omega}\,, \quad \beta(q,\theta') \equiv
b(q\sin\theta',q\cos\theta')\,,\nonumber
\end{eqnarray}\noindent
or, taking $u$ as the integration variable,
\begin{eqnarray}\label{fourier4}
\tilde{I}(\rho,\zeta,\omega) &=& \frac{me^{i\omega t_0}}{4\pi^4 r}
\int\limits_{\arcsin\zeta/r}^{\pi-\arcsin\zeta/r}
\frac{d\theta'\sin\theta'}{\displaystyle\sqrt{\sin^2\theta'-
\frac{\zeta^2}{r^2}}}\int\limits_{0}^{\infty}d q\, q
|\beta(q,\theta')|^2 \int\limits_{\sqrt{\bm{q}^2 + 2m\omega} -
|\bm{q}| }^{\sqrt{\bm{q}^2 + 2m\omega} + |\bm{q}|}d u \frac{\sin(u
r)}{u^2}\,.
\end{eqnarray}\noindent
To further transform this integral, it is convenient to define a
function $\Gamma(q,\theta,\omega)$ according to
\begin{eqnarray}\label{fourier5}
\Gamma(q,\theta) = \frac{1}{2\pi^2}\int\limits_{q}^{+\infty} dw\,
w |\beta(w,\theta)|^2\,.
\end{eqnarray}\noindent
Then integrating by parts, and taking into account that
$\Gamma(q,\theta,\omega)\to 0$ for $q \to \infty,$ brings
Eq.~(\ref{fourier4}) to the form
\begin{eqnarray}\label{fourier6}
\tilde{I}(\rho,\zeta,\omega) &=& \frac{me^{i\omega t_0}}{2\pi^2 r}
\int\limits_{\arcsin\zeta/r}^{\pi-\arcsin\zeta/r}
\frac{d\theta\sin\theta}{\displaystyle\sqrt{\sin^2\theta-
\frac{\zeta^2}{r^2}}} \int\limits_{0}^{+\infty}d q
\frac{\Gamma(q,\theta)}{\sqrt{q^2 + 2m\omega }}\nonumber\\&&
\times\biggl\{\left.\frac{\sin(u r)}{u}\right|_{u = \sqrt{q^2 +
2m\omega} + q} + \left.\frac{\sin(u r)}{u}\right|_{u = \sqrt{q^2 +
2m\omega} - q}\biggr\}\,.
\end{eqnarray}\noindent
Finally, integrating $\tilde{I}$ by parts once more, we find
\begin{eqnarray}
\tilde{I}(\rho,\zeta,\omega) &=& \frac{e^{i\omega t_0}}{2\pi^2
r\omega} \int\limits_{\arcsin\zeta/r}^{\pi-\arcsin\zeta/r}
\frac{d\theta\sin\theta}{\displaystyle\sqrt{\sin^2\theta-
\frac{\zeta^2}{r^2}}} \int\limits_{0}^{+\infty}d q \nonumber\\&&
\times\sin\left(r\sqrt{q^2 + 2m\omega }\right)
\left\{2\Gamma\cos(q r) +
\frac{1}{r}\frac{\partial\Gamma}{\partial q} \sin(q r) \right\}\,.
\label{i2}
\end{eqnarray}\noindent
This expression considerably simplifies in the practically
important case of low $\omega$ and large $r.$ Namely, if $\omega$
is such that
\begin{eqnarray}
\omega\ll \frac{\tilde{D}}{m r} \equiv \omega_0\,, \label{cond1}
\end{eqnarray}\noindent and also
\begin{eqnarray}
r\tilde{D} \gg 1 \label{cond2}
\end{eqnarray}\noindent
(and therefore, $\omega \ll \tilde{D}^2/m$), then the first term
in $\tilde{I}$ turns out to be exponentially small ($\sim e^{-r
\tilde{D}}$) because of the oscillating product of trigonometric
functions. Replacing $\sin^2(qr)$ by its average value (1/2) in
the rest of $\tilde{I}$ gives
\begin{eqnarray}
\tilde{I}(\rho,\zeta,\omega) &=& \frac{e^{i\omega t_0}}{4\pi^2
r^2\omega} \int\limits_{\arcsin\zeta/r}^{\pi-\arcsin\zeta/r}
\frac{d\theta\sin\theta}{\displaystyle\sqrt{\sin^2\theta-
\frac{\zeta^2}{r^2}}} \int\limits_{0}^{+\infty}d
q\frac{\partial\Gamma}{\partial q} = - \frac{e^{i\omega
t_0}}{4\pi^2 r^2\omega}
\int\limits_{\arcsin\zeta/r}^{\pi-\arcsin\zeta/r}
\frac{d\theta\sin\theta\Gamma(0,\theta)}{\displaystyle\sqrt{\sin^2\theta-
\frac{\zeta^2}{r^2}}}\,, \nonumber
\end{eqnarray}\noindent
or,
\begin{eqnarray}
\tilde{I}(\rho,\zeta,\omega) = - \frac{e^{i\omega t_0}}{8\pi^4
r^2\omega}\int\limits_{\chi}^{\pi-\chi}
\frac{d\theta\sin\theta}{\displaystyle\sqrt{\sin^2\theta-
\sin^2\chi}}\int\limits_{0}^{+\infty} d q\, q |\beta(q,\theta)|^2
\,, \label{i22}
\end{eqnarray}\noindent where $\chi$ is the angle between the vector
$\bm{r}$ and $x,y$ plane,  $\sin\chi = \zeta/r.$

In the case of a spherically symmetric wave function,
$\beta(q,\theta) = \beta(q),$ integration over $\theta$ in
Eq.~(\ref{i22}) yields the expression derived in
Ref.~\cite{kazakov}
\begin{eqnarray}
\tilde{I}(\rho,\zeta,\omega) = - \frac{e^{i\omega t_0}}{8\pi^3
r^2\omega}\int\limits_{0}^{+\infty} d q\, q |\beta(q)|^2 \,.
\label{i23}
\end{eqnarray}\noindent
It has been assumed in the course of derivation of Eq.~(\ref{i22})
that $\omega>0,$ $\zeta >0.$ It is not difficult to verify that in
the general case, $\tilde{I}(\rho,\zeta,\omega) = e^{i(\omega -
|\omega|)t_0}\tilde{I}(\rho,|\zeta|,|\omega|).$

Substituting the obtained expression into the defining equations
(\ref{meanfourier}), (\ref{corr}), (\ref{corrf}), we thus obtain
the following expression for the low-frequency asymptotic of the
correlation function
\begin{eqnarray}\label{main1}
C_{00}(\rho,\zeta,r',\omega) &=& e^{i\omega (t_0 - t')
}\frac{e^2}{32\pi^5 r^2 r'
|\omega|}\int\limits_{|\chi|}^{\pi-|\chi|}
\frac{d\theta\sin\theta}{\displaystyle\sqrt{\sin^2\theta-
\sin^2\chi}}\int\limits_{0}^{+\infty} d q\,
q|\beta(q,\theta)|^2\,,
\end{eqnarray}\noindent all other components of the correlation
function being suppressed by the factor $|\bm{q}|/m.$

In applications to microelectronics, $\omega$ varies from $10^{-6}
{\rm Hz}$ to $10^6\,{\rm Hz},$ the relevant distances $r$ are
usually $10^{-5}\,{\rm cm}$ to $10^{-2}\,{\rm cm},$ $\tilde{D}\sim
\hbar/d,$ where $d\approx 10^{-8}\,{\rm cm}$ is the lattice
spacing, and $m$ is the effective electron mass, $m\approx
10^{-27}\,{\rm g},$ hence, $\omega_0 \approx 10^{10}\,{\rm Hz},$
so the conditions $r\gg d,$ $\omega\ll\omega_0$ are always
well-satisfied.

In connection with the application of the obtained results to
solids, it should be stressed that they refer to long-living
free-evolving electron states. At the same time, because of
collisions of electrons with phonons, impurities, and with each
other, their evolution in a crystal usually cannot be considered
free. It is important, however, that in view of smallness of the
electron mass in comparison with the atomic masses, the electron
collisions with phonons and impurities may often be considered
elastic. Such collisions do not change the electron energy, and
therefore they do not influence time evolution of the electron
wave function. As to the electron-electron collisions, they do
change the energy of electrons. However, the electron component in
solids is practically always degenerate, and hence, only electrons
with energies near the Fermi surface are actually scattered.
Therefore, the above results concerning dispersion of the
electromagnetic field fluctuations remain essentially the same
despite the electron collisions, provided that they are applied to
electrons far from the Fermi surface, and expressed in terms of
the electron density matrix, rather than the wave function.
Denoting the diagonal elements of the momentum space density
matrix of electrons by $\varrho(q,\theta),$ Eq.~(\ref{main1}) thus
takes the form
\begin{eqnarray}\label{main2}
C_{00}(\rho,\zeta,r',\omega) &=& e^{i\omega (t_0 - t')
}\frac{e^2}{32\pi^5 r^2 r'
|\omega|}\int\limits_{|\chi|}^{\pi-|\chi|}
\frac{d\theta\sin\theta}{\displaystyle\sqrt{\sin^2\theta-
\sin^2\chi}}\int\limits_{0}^{+\infty} d q\, q\varrho(q,\theta)\,.
\end{eqnarray}\noindent The power spectrum
of the noise produced by uncorrelated electrons in a sample can be
found by integrating Eq.~(\ref{main2}) with respect to $\bm{x}_0$
over the sample volume. If the time instants $t_0$ are distributed
uniformly (which is natural to expect), then the value of the
total noise spectrum function, $C^{\rm tot}_{00},$ remains at the
level of the individual contribution (\ref{main2}) independently
of the number of electrons in the sample, because of cancellation
of the alternating phase factors $e^{i\omega t_0}$.

\section{The noise induced by external electric field}\label{external}

According to the property 3) of observed flicker noise, mentioned
in the introduction, power spectrum of the noise induced by
external electric field is proportional to $\bm{E}^2,$ at least
for sufficiently small values of the field strength $\bm{E}.$ If
the density matrix $\varrho$ were analytic with respect to
$\bm{E},$ then the scalar function $C^{\rm tot}_{00}(\bm{E}) -
C^{\rm tot}_{00}(0)\equiv \Delta C^{\rm tot}_{00}(\bm{E})$ would
expand in even powers of $\bm{E},$ and therefore, the leading term
would be quadratic in the field strength, as required. However,
$\varrho$ is not generally analytic in $\bm{E},$ and therefore,
the expansion of $\Delta C^{\rm tot}_{00}(\bm{E})$ might contain,
e.g., a term proportional to $|\bm{E}|,$ in contradiction with the
experiment. Thus, our primary concern below will be the weak field
asymptotic of the power spectrum given by Eq.~(\ref{main2}). There
is actually a simple and general reason why the right hand side of
this equation should be quadratic in $\bm{E}$ despite possible
non-analyticity of the density matrix $\varrho.$ This matrix is a
functional of the equilibrium density matrix, $\varrho_0,$ and of
the field strength. As was mentioned in the preceding section, the
electron states responsible for the $1/\omega$ behavior of the
correlation function are those with energies far from the Fermi
surface. For such states, the function $\varrho$ is a constant
inversely proportional to the electron density, in particular, it
is independent of parameters characterizing the Fermi surface,
such as Fermi energy or momentum. Therefore, on dimensional
grounds, the integral on the right of Eq.~(\ref{main2}) should be
proportional to the square of a characteristic momentum built from
the field strength. The only such momentum is $e|\bm{E}|\tau,$
where $\tau$ is some time parameter characterizing electron
kinetics, and hence, $\Delta C^{\rm tot}_{00}\sim \bm{E}^2\,.$
This reasoning will be illustrated below using the simplest
kinetic model with the relaxation-type collision term, which
admits full theoretic investigation.

In the presence of a constant homogeneous electric field, the
model kinetic equation reads
\begin{eqnarray}\label{keq}
e E \frac{\partial\varrho}{\partial q_z} = - \frac{\varrho -
\varrho_0}{\tau}\,,
\end{eqnarray}\noindent
where $z$ axis is chosen in the direction of $\bm{E}$ ($E_z =
|\bm{E}|\equiv E$), $\tau$ is the relaxation time, and
$\varepsilon$ the electron energy. Since Eq.~(\ref{main2}) was
obtained for a free electron, we assume that the band structure in
the given solid is parabolic, $\varepsilon = \bm{q}^2/2m^*,$ where
$m^*$ is the effective electron mass. According to
Eq.~(\ref{norm}), the function $\varrho$ is normalized by
$$\int \frac{d^3\bm{q}}{(2\pi)^3}\varrho(\varepsilon) = 1\,.$$
If the function $\varrho_0$ satisfies this condition, then the
normalized solution of the kinetic equation (\ref{keq}) has the
form
\begin{eqnarray}\label{rho}
\varrho = \frac{e^{q_z/q_0}}{q_0}\int\limits_{q_z}^{\infty}d\xi
e^{-\xi/q_0}\varrho_0(\varepsilon')\,, \quad \varepsilon' =
\frac{1}{2m^*}(q^2_{\bot} + \xi^2)\,, \quad q_0 = |e|E\tau\,.
\end{eqnarray}\noindent Thus, in order to find the quantity
$C_{00},$ we have to calculate the following integral
\begin{eqnarray}
K = \int\limits_{\chi}^{\pi-\chi}
\frac{d\theta\sin\theta}{\displaystyle\sqrt{\sin^2\theta-
\sin^2\chi}}\int\limits_{0}^{+\infty} d q\, q
\frac{e^{q\cos\theta/q_0}}{q_0}
\int\limits_{q\cos\theta}^{\infty}d\xi
e^{-\xi/q_0}\varrho_0(\varepsilon')\,, \quad \varepsilon' =
\frac{1}{2m^*}(q^2\sin^2\theta + \xi^2)\,. \nonumber
\end{eqnarray}\noindent Changing the integration variables
$q,\theta\to \zeta,\eta$ according to $$\cos\theta =
\frac{\zeta\cos\chi}{\sqrt{\zeta^2 + \eta^2}}\,, \quad q^2 =
\zeta^2 + \eta^2$$ brings this integral to the form
\begin{eqnarray}
K = \iint\limits_{-\infty}^{+\infty}d\zeta d\eta
\frac{e^{\zeta\cos\chi/q_0}}{2q_0}
\int\limits_{\zeta\cos\chi}^{\infty}d\xi
e^{-\xi/q_0}\varrho_0(\varepsilon'')\,, \quad \varepsilon'' =
\frac{1}{2m^*}(\eta^2 + \zeta^2\sin^2\chi + \xi^2)\,. \nonumber
\end{eqnarray}\noindent Integrating by parts with respect to $\zeta$ yields
\begin{eqnarray}
K = \frac{1}{2}\iint\limits_{-\infty}^{+\infty} d\zeta d\eta
\varrho_0\left(\frac{\zeta^2 + \eta^2}{2m^*}\right) -
\frac{\sin^2\chi}{2m^*\cos\chi}\iint\limits_{-\infty}^{+\infty}d\zeta
d\eta~\zeta e^{\zeta\cos\chi/q_0}
\int\limits_{\zeta\cos\chi}^{\infty}d\xi e^{-\xi/q_0}
\frac{d\varrho_0}{d\varepsilon}(\varepsilon'') \,. \nonumber
\end{eqnarray}\noindent The first term in this expression
represents the value of the quantity $K$ in the absence of the
external field. Therefore, performing a shift $\xi \to \xi +
\zeta\cos\chi,$ and then $\zeta \to \zeta - \xi\cos\chi$ in the
remaining integral, we find the part induced by the electric field
\begin{eqnarray}
\Delta K = \frac{\sin^2\chi}{2m^*}\iint
\limits_{-\infty}^{+\infty} d\zeta d\eta
\int\limits_{0}^{\infty}d\xi~\xi
e^{-\xi/q_0}\frac{d\varrho_0}{d\varepsilon}(\varepsilon''')\,,
\quad \varepsilon''' = \frac{1}{2m^*}(\eta^2 + \zeta^2 + \xi^2
\sin^2\chi)\,. \nonumber
\end{eqnarray}\noindent
Going over to the polar coordinates in the $(\zeta,\eta)$ plane
gives finally
\begin{eqnarray}
\Delta K = \pi\sin^2\chi\int\limits_{0}^{\infty}d\xi~\xi
e^{-\xi/q_0}\int\limits_{\xi^2\sin^2\chi/2m^*}^{+\infty}
d\varepsilon \frac{d\varrho_0}{d\varepsilon}(\varepsilon) = -
\pi\sin^2\chi\int\limits_{0}^{\infty}d\xi~\xi
e^{-\xi/q_0}\varrho_0\left(\frac{\xi^2\sin^2\chi}{2m^*}\right)\,.
\nonumber
\end{eqnarray}\noindent In the case of a degenerate
electron system, the function $\varrho_0(\varepsilon)$ is nearly
constant up to Fermi energy where it falls off to zero. For such a
function, $\Delta K \sim q_0^2,$ up to exponentially small terms,
provided that $q_0$ is sufficiently small. Indeed, assuming $q_0
\ll q_F,$ where $q_F$ is the electron momentum at the Fermi
surface, and neglecting terms of the order $e^{- q_F/q_0},$ one
can substitute
$\varrho_0\left(\frac{\xi^2\sin^2\chi}{2m^*}\right)$ by
$\varrho_0(0)$ in the above integral to obtain
\begin{eqnarray}
\Delta K = -
\pi\sin^2\chi\varrho_0(0)\int\limits_{0}^{\infty}d\xi~\xi
e^{-\xi/q_0} = - q_0^2\varrho_0(0)\pi\sin^2\chi\,. \nonumber
\end{eqnarray}\noindent Inserting this into Eq.~(\ref{main2}),
we arrive at the following expression for the power spectrum of
the induced noise
\begin{eqnarray}\label{main3}
\Delta C_{00} &=& - e^{i\omega (t_0 - t') }\frac{e^2
q_0^2\varrho_0(0)\zeta^2}{32\pi^4 r^4 r' |\omega|}\,.
\end{eqnarray}\noindent Thus, $\Delta C_{00} \sim \bm{E}^2,$ as
was to be shown. It is important that the correction terms are of
the order $\exp^{-q_F/q_0},$ and hence, from the practical point
of view, the condition $q_0 \ll q_F$ amounts to $q_0 < q_F.$ This
implies that the upper limit on the field strengths for which
Eq.~(\ref{main3}) is valid is well above all experimentally
relevant values.

Let us compare this result with the power spectrum of quantum
fluctuations in the absence of external field, obtained in
Ref.~\cite{kazakov}, which in the present notation has the form
\begin{eqnarray}
C_{00}(0) = e^{i\omega (t_0 - t') }\frac{e^2}{32\pi^4 r^2 r'
|\omega|}\int\limits_{0}^{+\infty} d q\, q \varrho_0(\varepsilon)
\,.\nonumber
\end{eqnarray}\noindent By the order of magnitude,
$$\frac{\Delta C_{00}}{C_{00}(0)} \sim \frac{q^2_0\varrho_0(0)}{
\int\limits_{0}^{+\infty} d q\, q\varrho_0} \sim q_F
q^2_0\varrho_0(0) \sim \frac{q^2_0}{q^2_F}\,,$$ since
$\varrho_0(0) \sim q^{-3}_F\,.$ Thus, in the model considered, the
induced noise represents a relatively small correction.

In practice, one is interested in fluctuations of the voltage,
$U,$ between two leads attached to a sample. Using the above
results, it is not difficult to write down an expression for the
voltage correlation function, $C_{\rm U}(\bm{x},\bm{x}',\omega)$.
We have
\begin{eqnarray}\label{comment}
\langle \hat{U}(\bm{x},\bm{x}',t)\rangle \langle
\hat{U}(\bm{x},\bm{x}',t')\rangle &=& \langle \hat{A}_0(\bm{x},t)
- \hat{A}_0(\bm{x}',t)\rangle \langle \hat{A}_0(\bm{x},t') -
\hat{A}_0(\bm{x}',t')\rangle \nonumber\\ &=& \langle
\hat{A}_0(\bm{x},t)\rangle \langle \hat{A}_0(\bm{x},t')\rangle +
\langle
\hat{A}_0(\bm{x}',t)\rangle \langle \hat{A}_0(\bm{x}',t')\rangle \nonumber\\
&-& \langle \hat{A}_0(\bm{x}',t)\rangle \langle
\hat{A}_0(\bm{x},t')\rangle - \langle \hat{A}_0(\bm{x},t)\rangle
\langle \hat{A}_0(\bm{x}',t')\rangle\,,
\end{eqnarray}\noindent where $\langle \cdots \rangle$ denotes
averaging over the given in state. Fourier transforming and
applying Eq.~(\ref{main3}) to each of the four terms in this
expression, we obtain the power spectrum of the induced voltage
fluctuation across the sample
\begin{eqnarray}\label{main4}
\Delta C_{U} &=& - e^{i\omega (t_0 - t') }\frac{e^2
q_0^2\varrho_0(0)}{32\pi^4 |\omega|}\left(\frac{\zeta^2}{r^5} +
\frac{\zeta'^2}{r'^5} - \frac{\zeta'^2}{r'^4 r} -
\frac{\zeta^2}{r^4 r'}\right)\,.
\end{eqnarray}\noindent
This equation represents an individual contribution of an electron
to the electric potential fluctuation. As was mentioned at the end
of Sec.~\ref{discon}, because of the oscillating exponent
$e^{i\omega t_0},$ the magnitude of the total noise remains at the
level of the individual contribution independently of the number
of electrons. Therefore, summing up all contributions amounts
simply to averaging over $\bm{x}_0:$
\begin{eqnarray}\label{main5}
\left|\Delta C_{U}^{\rm tot}\right| = \frac{e^2
q_0^2\varrho_0(0)G}{32\pi^4|\omega|\Omega}\,,
\end{eqnarray}\noindent
where
\begin{eqnarray}\label{gfactor}
G \equiv \int\limits_\Omega d^3\bm{x}_0\left(\frac{\zeta^2}{r^5} +
\frac{\zeta'^2}{r'^5} - \frac{\zeta'^2}{r'^4 r} -
\frac{\zeta^2}{r^4 r'}\right)
\end{eqnarray}\noindent
is a dimensionless geometrical factor, and $\Omega$ is the sample
volume.

It is convenient to express the right hand side of
Eq.~(\ref{main5}) through the number of electrons in the sample,
$N,$ rather than the sample volume. Taking into account that for
degenerate electrons, $\varrho_0(0) = 2/n,$ where $n$ is the
electron density (the factor $2$ accounts for two spin states),
substituting $n = N/\Omega,$ $q_0 = |e|E\tau,$ and restoring the
ordinary units brings Eq.~(\ref{main5}) to the form
\begin{eqnarray}\label{main6}
\left|\Delta C_{U}^{\rm tot}\right| = \frac{e^4}{32\pi^5
\hbar^2}\frac{E^2 \tau^2 G}{N f}\,,
\end{eqnarray}\noindent where $f = |\omega|/2\pi.$ Finally, if the
voltage leads are aligned in $z$ direction, as is usually the
case, Eq.~(\ref{main6}) can be rewritten as
\begin{eqnarray}\label{main7}
\left|\Delta C_{U}^{\rm tot}\right| = \frac{\alpha \bar{U}^2}{N
f}\,, \quad \alpha = \frac{e^4 \tau^2 G}{32\pi^5 l^2 \hbar^2}\,,
\end{eqnarray}\noindent where $\bar{U}$ is the average voltage
across the sample, and $l$ is the sample length in the direction
of the electric field. In this form, it is similar to the
well-known empirical Hooge law, except that $N$ in
Eq.~(\ref{main7}) is the total number of electrons in the sample,
rather than the number of charge carriers. We see that the analog
of the Hooge constant, $\alpha,$ depends on physical properties of
the sample material as well as on the sample geometry.

\begin{figure}
\includegraphics{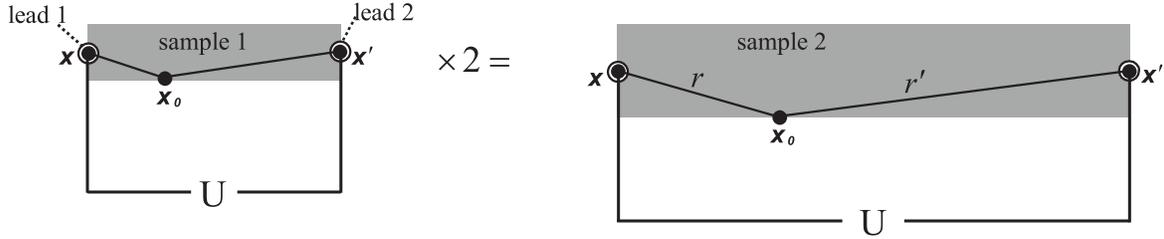}
\caption{Voltage measurement in geometrically similar samples
($s=2$).} \label{fig3}
\end{figure}

Let us discuss the role of the sample geometry in somewhat more
detail. Consider two geometrically similar samples, and let $s$ be
the ratio of their linear dimensions (see Fig.~\ref{fig3}). It
turns out that such samples are characterized by the same value of
the $G$-factor. Indeed, since the voltage across the sample is
usually measured via two leads attached to its surface, the
radius-vectors of the leads drawn from the center of similitude
scale by the same factor $s,$ and therefore,
\begin{eqnarray}
\int\limits_{\Omega_2}d^3\bm{x}_0\frac{\zeta^2}{r^4 r'} &=&
\int\limits_{s^3\Omega_1}d^3\bm{x}_0\frac{(s z - z_0)^2}{|s\bm{x}
- \bm{x}_0|^4 |s\bm{x}' - \bm{x}_0|} = \int\limits_{\Omega_1}
s^3d^3\bm{x}_0\frac{(s z - s z_0)^2}{|s\bm{x} - s\bm{x}_0|^4
|s\bm{x}' - s\bm{x}_0|} \nonumber\\ &=& \int\limits_{\Omega_1}
d^3\bm{x}_0\frac{(z - z_0)^2}{|\bm{x} - \bm{x}_0|^4 |\bm{x}' -
\bm{x}_0|} = \int\limits_{\Omega_1} d^3\bm{x}_0\frac{\zeta^2}{r^4
r'}\,,\nonumber
\end{eqnarray}\noindent
and likewise for the other terms in Eq.~(\ref{gfactor}). Thus, it
follows from Eq.~(\ref{main5}) that for geometrically similar
samples, the noise level is inversely proportional to the sample
volume. This is in agreement with the property 1) of flicker
noise, mentioned in Sec.~\ref{introduction}. It is also clear that
the distribution of the noise magnitude around the value
$C_{00}^{\rm tot}$ is Gaussian, by virtue of the central limiting
theorem. Thus, the quantum field fluctuations in a sample possess
the property 2) as well.

It is worth also to make the following comment concerning
expression (\ref{gfactor}). The integrand in this formula involves
the terms $\zeta^2/r^5$ and $\zeta'^2/r'^5$ which give rise
formally to a logarithmic divergence when the observation points
approach the sample. In this connection, it should be recalled
that the above calculations have been carried out under the
condition $r\tilde{D}\gg 1,$ hence, $r,r'$ cannot be taken too
small. Furthermore, one should remember that in any field
measurement in a given point, one deals actually with the field
averaged over a small but finite domain surrounding this point,
i.e., the voltage lead in our case. Thus, for instance, the
quantity $\hat{A}_0(\bm{x},t)$ appearing in the expression
(\ref{comment}) is to be substituted by $$\hat{\EuScript{A}}_0 =
\frac{1}{\Omega_{\rm L}}\int\limits_{\Omega_{\rm L}}d^3
\bm{x}\hat{A}_0(\bm{x},t)\,,$$ where $\Omega_{\rm L}$ is the
voltage lead volume. As a result of this substitution, the term
$\zeta^2/r^5,$ for instance, takes the form
$$\frac{1}{\Omega^2_{\rm L}}\iint\limits_{\Omega_{\rm L}}d^3
\bm{x} d^3\tilde{\bm{x}}\frac{(z - z_0)^2}{|\bm{x} - \bm{x}_0|^4
|\tilde{\bm{x}} - \bm{x}_0|} \,.$$ Upon substituting into
Eq.~(\ref{gfactor}), this term gives rise to a finite contribution
even for intersecting $\Omega, \Omega_L.$

\section{Conclusions}\label{conclude}

The main result of the present work is the general formula
(\ref{main1}) describing the power spectrum of quantum
electromagnetic fluctuations produced by elementary particles in
the presence of an external field. This formula shows that in the
low-frequency limit, the power spectrum exhibits an inverse
frequency dependence. Although the range of applicability of
Eq.~(\ref{main1}), given by the conditions (\ref{cond1}),
(\ref{cond2}), depends on the problem under consideration, it
embraces virtually all experimentally relevant frequencies and
distances. In particular, in application to microelectronics, the
term ``low-frequency limit'' means that $f \ll 10^{10}\,{\rm Hz},$
which covers well the whole measured band. To the best of the
author knowledge, none of the other physical mechanisms of flicker
noise, suggested so far, has been able to explain the observed
plenum of the $1/f$ law.

We have applied the general formula to the calculation of the
power spectrum of fluctuations produced by electrons in a sample
in external electric field, assuming that the electron kinetics is
described by a model equation with the relaxation-type collision
term. This calculation shows that the power spectrum of induced
fluctuations is proportional to the field strength squared for all
practically relevant values of the electric field. We have also
argued that this conclusion in fact holds true in the general
case. Finally, we have established the exact dependence of the
power spectrum on the sample geometry. We have shown that for
geometrically similar samples the noise power spectrum is
inversely proportional to the sample volume, while for samples of
the same volume dependence on the sample geometry is described by
the dimensionless $G$-factor given by Eq.~(\ref{gfactor}).

Qualitatively, the established properties of quantum
electromagnetic fluctuations match perfectly with the
experimentally observed properties of $1/f$-noise, and suggest
that these fluctuations can be considered as one of the underlying
mechanisms of flicker noise. As to the quantitative side, the
estimates of Ref.~\cite{kazakov} show that the noise level
predicted by Eq.~(\ref{main1}) in the absence of external field is
in a reasonable agreement with experimental data. At the same
time, according to the results of Sec.~\ref{external}, the change
of the noise level in external electric field is relatively small,
which disagrees with observations. Of course, it is difficult to
expect that predictions of the simple model employed in
Sec.~\ref{external} will be quantitatively correct. Although the
found disagreement may be the result of the model
oversimplification, it raises the question of possible alternative
mechanisms of the noise amplification by external field. Recall
that the expression (\ref{main5}) for the power spectrum of
fluctuations produced by a sample was derived assuming uniform
distribution of the time instants $t_0.$ Therefore, correlation
between $t_0$'s is a possible source of the noise amplification.
Since the number of electrons in the sample is large, already a
relatively small correlation in the values of $t_0$ would result
in a noticeable increase of the noise level. Thus, the question
concerning the relative role of the quantum electromagnetic
fluctuations in explaining the observed $1/f$ noise requires
further investigation.

\acknowledgments{I thank Drs. G.~A.~Sardanashvili,
K.~V.~Stepanyantz, and especially P.~I.~Pronin (Moscow State
University) for interesting discussions.}

\pagebreak

\end{document}